\documentclass{article}
\usepackage{spconf,amsmath,amssymb,graphicx,booktabs,tabularx,array,microtype,balance}
\usepackage[hidelinks]{hyperref}
\hypersetup{pdftitle={Robust Cross-Device Speaker Linking through Spatiotemporal Structure},pdfauthor={Suhas BN and Nikitha K}}
\newcolumntype{Y}{>{\raggedright\arraybackslash}X}
\title{UNSUPERVISED CROSS-DEVICE SPEAKER DIARIZATION AND LINKING THROUGH SPATIOTEMPORAL STRUCTURE}
\name{Suhas BN, Nikitha K}
\address{Independent Researchers}
\begin{document}
\maketitle

\begin{abstract}
Linking speakers across independently recorded rooms is difficult because microphone, room, and codec mismatch distort speaker embeddings. We study visitor linking without visitor enrollment or identity labels in simulated outpatient workflows. The proposed method combines per-device embedding centering, concurrency constraints, room-type-specific visit constraints, and a clustering threshold calibrated from concurrent visitors and identified staff voices. On simulated days built from VOiCES recordings, with companions, return visits, walk-ins, silent visits, and acoustic leakage, it improves B$^3$-F1 over the best label-free baseline for all 11 extractors, by 5.3-14.6 points (all 95\% intervals exclude zero). ECAPA-TDNN reaches 86.6, compared with 76.4 for the best label-free baseline and 87.7 for the best oracle-informed baseline. Gains persist across the reported workflow perturbations, but depend on removing duplicated cross-room speech: without this step, the method wins for only 2 of 11 extractors. Single-visit patients also remain vulnerable to false merges. These results support structure-aware linking while identifying preprocessing and identity-safety limitations that must be addressed before clinical deployment.
\end{abstract}

\begin{figure}[t]
\centering
\includegraphics[width=\columnwidth]{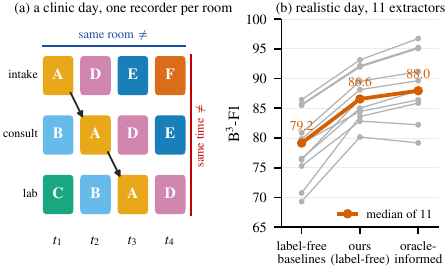}
\caption{(a) Illustrative visit schedule: patient A moves between rooms. Concurrent visits cannot share an identity after cross-room duplicate removal; same-room exclusions apply only where repeat visits are disallowed. (b) Default simulated day: gray lines show the 11 extractors and orange shows the median B$^3$-F1. Oracle-informed baselines use privileged identity information.}
\label{fig:method}
\end{figure}

\vspace{-0.1cm}
\section{Introduction}
\label{sec:intro}
\vspace{-0.1cm}
People in outpatient facilities move between reception, consultation, laboratory, and pharmacy. Linking their speech across room recordings could support documentation and workflow analysis without requiring visitor enrollment. This task differs from diarization within one recording: the same speaker is observed through different microphones, reverberation, noise, and codecs. Such mismatch is a known challenge in far-field speaker verification \cite{martin2010nist,qin2020ffsvc,nandwana2019voices,zheng2023threedspeaker} and can cause clustering to reflect recorder characteristics rather than speaker identity.

A facility's schedule supplies information beyond the acoustic embeddings. A person cannot occupy two rooms simultaneously. Some room types also admit at most one visit per person, although this rule fails wherever patients return. Finally, identified staff members who speak on multiple recorders provide cross-device calibration examples. These signals can constrain linking without visitor identity labels, but their usefulness depends on how faithfully the observed audio represents the underlying visits. We combine per-device centering with schedule-derived constraints and staff-based threshold calibration. We evaluate the method on simulated clinic days containing companions, return visits, walk-ins, silent visits, cross-room leakage, and waiting-room television. The evaluation covers 11 extractors, from a 1.1M-parameter network to a 587M-parameter self-supervised model. 

Our contributions are: (1) a room-type-aware linking formulation, (2) comparison with label-free and oracle-informed baselines across these extractors, and (3) a sensitivity analysis exposing both gains and failure modes. In particular, the method improves on label-free baselines across models, but neither duplicated speech nor singleton identity errors are resolved by schedule constraints alone.

\vspace{-0.1cm}
\section{Related Work}
\label{sec:related}
\vspace{-0.2cm}
\textbf{Speaker representations and channel mismatch.} NIST and far-field evaluations document the effects of channel and acoustic mismatch \cite{martin2010nist,qin2020ffsvc,nandwana2019voices}; 3D-Speaker broadens multi-device evaluation \cite{zheng2023threedspeaker}. Following x-vectors \cite{snyder2018xvector}, architectures including ECAPA-TDNN, TitaNet, ERes2Net, and ReDimNet improve speaker embeddings \cite{desplanques2020ecapa,koluguri2022titanet,chen2023eres2net,yakovlev2024redimnet,yakovlev2026redimnet2}. ReDimNet2+ targets channel and codec robustness \cite{redimnet2plus}, while w2v-BERT systems exploit self-supervised pretraining \cite{li2025w2vbert}. Speech-aware LLMs do not automatically provide reliable speaker discrimination \cite{thebaud2026speechllm}. We instead study workflow information as a complement to frozen speaker extractors.

\textbf{Linking and constrained clustering.} Cross-recording diarization associates identities across recordings \cite{lelan2018adaptive}. Adaptation and calibration with unlabeled data address condition mismatch \cite{castan2017}, including adaptation of PLDA through CORAL+ \cite{lee2019coral}. Within-recording diarization incorporates constraints from neural segmentation or conversational roles \cite{kinoshita2021integrating,flemotomos2022roles}. Spatiotemporal constraints also guide tracklet association in visual person re-identification \cite{xie2020progressive}. Our contribution is the combination of device centering, room-specific constraints, and staff-based calibration for cross-room speaker linking.

\textbf{Audio front ends and deployment.} Multi-room speech detection and localization have been studied with microphone arrays \cite{giannoulis2014athena}; separating live conversation from television is a distinct recognition problem \cite{georgiou2023tv}. Toolkits such as SpeechBrain and pyannote.audio support speech modeling and diarization \cite{ravanelli2021speechbrain,bredin2023pyannote}, but do not establish the accuracy of the upstream segments assumed here. Low-bit quantization can also change verification decisions \cite{quant2026sv}; parameter count alone therefore does not demonstrate reliable on-device operation.

\begin{table*}[t]
\centering
\caption{Default simulated day: 40 patients, four rooms, hard recorder mix; 20 days. LF: best label-free baseline; Oracle: best oracle-informed baseline; Itin.: exact itinerary recovery; SM: single-visit patients wrongly merged. Patient-level columns refer to the proposed method. Scores are scaled by 100; arrows indicate the preferred direction. All gains significant (95\% $t$-intervals).}
\vspace{0.2cm}
\label{tab:models}
\fontsize{8.5}{10}\selectfont
\setlength{\tabcolsep}{3pt}
\renewcommand{\arraystretch}{1.10}
\begin{tabularx}{\textwidth}{@{}llrYrrrrr@{}}
\toprule
 & & & & \multicolumn{3}{c}{B$^3$-F1 $\uparrow$} & \multicolumn{2}{c}{Patients (\%)} \\
Family & Model & Params & Training data & LF & \textbf{Ours} & Oracle & Itin.$\uparrow$ & SM$\downarrow$ \\
\midrule
TDNN & ECAPA-TDNN & 20.8M & VoxCeleb1+2 & 76.4 & \textbf{86.6} & 87.7 & 57 & 57 \\
Depthwise 1D & TitaNet-S$^\dagger$ & 6.4M & Mixed$^\ddagger$ & 70.7 & \textbf{83.6} & 85.9 & 49 & 65 \\
 & TitaNet-L$^\dagger$ & 25.3M & Mixed$^\ddagger$ & 65.6 & \textbf{80.2} & 79.2 & 41 & 72 \\
Res2Net & ERes2Net & 6.6M & VoxCeleb & 76.5 & \textbf{85.0} & 88.0 & 50 & 65 \\
 & ERes2Net-base & 9.9M & 3D-Speaker & 75.3 & \textbf{82.8} & 82.2 & 47 & 60 \\
Hybrid & ReDimNet2-B0 & 1.1M & VoxCeleb2 & 79.2 & \textbf{84.5} & 86.4 & 51 & 64 \\
 & ReDimNet2-B3 & 4.2M & VoxCeleb2 & 79.8 & \textbf{88.1} & 89.7 & 64 & 48 \\
 & ReDimNet2-B3 & 4.5M & VB2+Vox2+CNC2 & 85.6 & \textbf{92.1} & 95.2 & 73 & 43 \\
 & ReDimNet2-B6 & 12.4M & VoxCeleb2 & 80.9 & \textbf{89.6} & 91.2 & 66 & 53 \\
 & ReDimNet2+ & 12.4M & Multi-corpus + codecs & 85.5 & \textbf{92.0} & 95.1 & 73 & 39 \\
SSL & w2v-BERT 2.0 SV & 587M & SSL 4.5M h + Vox2 & 86.4 & \textbf{93.1} & 96.7 & 76 & 35 \\
\bottomrule
\end{tabularx}
\par\smallskip\raggedright $^\dagger$LibriSpeech training may overlap evaluation speakers. $^\ddagger$VoxCeleb, Fisher, Switchboard, LibriSpeech, and SRE. Vox1/2 denote VoxCeleb1/2; SSL denotes self-supervised pretraining.
\end{table*}

\vspace{-0.2cm}
\section{Method}
\label{sec:method}
\vspace{-0.2cm}
\subsection{Observations and assumptions}
\vspace{-0.1cm}
A facility has $R$ rooms, each with a recorder. Each visitor observation $v$ has a room $r(v)$, a time slot $t(v)$, and $k_v$ speech chunks with unit-normalized embeddings $x_{v,i}$. We assume within-recording processing supplies visitor speech separately from staff. Staff identities are established once; at least one staff member must be observed on multiple recorders for positive calibration. Thus, ``label-free'' refers to visitor linking, not to staff identification or extractor training.

The default system is an offline day-level linker. Independently captured recordings still require a common time-slot reference for concurrency constraints. Cross-room duplicate removal is needed because the same utterance may be audible on several recorders. Media rejection reduces contributions from television. These preprocessing components are part of the evaluated pipeline and their removal is discussed separately in Section~\ref{sec:results}.

\vspace{-0.1cm}
\subsection{Device centering}
\vspace{-0.1cm}
Let $\mu_d$ denote the mean embedding computed for recorder $d$ over the day. We subtract this recorder-specific mean before forming each normalized visitor centroid:
\vspace{-0.05cm}
\begin{equation}
 z_v=\sum_{i=1}^{k_v}(x_{v,i}-\mu_{d(v)}),\qquad
 c_v=\frac{z_v}{\lVert z_v\rVert_2}.
\label{eq:center}
\end{equation}
Cosine similarity between visits is $s(u,v)=c_u^\top c_v$. Centering reduces shared recorder offsets without requiring labeled visitor adaptation data. It does not eliminate all channel effects, and its reliability depends on the speech contributing to the mean.

\vspace{-0.3cm}
\subsection{Room-type-aware constraints}
\vspace{-0.1cm}
The cannot-link set $\mathcal C$ combines two rules. \textit{Concurrency:} observations in different rooms at the same time are assigned different identities, provided they are distinct visits rather than duplicated audio. \textit{Room uniqueness:} observations in the same room are assigned different identities only when that room's protocol rules out repeat visits. Let $\mathcal R_{\mathrm{once}}$ be this subset of rooms. Then
\begin{equation}
\begin{split}
\mathcal C={}&\{(u,v):t(u)=t(v),\ r(u)\ne r(v)\}\\
 &{}\cup\{(u,v):u\ne v,\ r(u)=r(v)\in\mathcal R_{\mathrm{once}}\}.
\end{split}
\label{eq:constraints}
\end{equation}
Allowing returns in other room types avoids treating a workflow assumption as a universal identity rule. Clock errors and residual acoustic leakage can invalidate even the concurrency constraint.

\vspace{-0.1cm}
\subsection{Calibration and linking}
\vspace{-0.1cm}
Concurrent visitor pairs supply negative cosine scores. Positive scores come from the same identified staff member on different recorders, using disjoint speech content and chunk counts matched to visitor observations. The stopping threshold $\tau$ is selected at the empirical equal-error operating point between these score distributions. The concurrency-only variant instead uses the 99th percentile of negative scores, as in the evaluated configuration.

Average-linkage agglomerative clustering (AHC) starts with one cluster per observation. At each step it merges the admissible pair with greatest average cross-cluster cosine similarity, provided that similarity is at least $\tau$. A merge is inadmissible if any constituent pair belongs to $\mathcal C$. This produces speaker clusters without specifying the number of visitors. The results also include variants using the preceding day's statistics, a leakage guard, and causal greedy linking; these are distinct from the default offline procedure.

\section{Experimental Setup}
\label{sec:setup}
\vspace{-0.2cm}
\subsection{Recordings and simulated workflows}

The experiments use VOiCES \cite{richey2018voices}, which re-records LibriSpeech speech \cite{panayotov2015librispeech} in real rooms. We select 100 test speakers, balanced by sex, with near and far microphone recordings in quiet and babble. Phone-like processing adds automatic gain control and Opus at 12\,kb/s, or 8\,kHz GSM~06.10. The hard recorder mix comprises Opus, a near microphone with babble, a far microphone, and GSM; the easy mix uses Opus and three quiet studio microphones.

The default day has 40 patients, four rooms, four staff members, and 6\,s of speech per speaking visit, assembled from 3\,s chunks. Each setting is evaluated over 20 simulated days. Speech content is disjoint across a person's visits and across staff calibration recordings. The realistic workflow includes companions (70\%), laboratory-related returns (85\% of laboratory patients), silent visits (20\%), cross-room leakage (15\%), walk-ins (15\%), and television in 50\% of slots. Table~\ref{tab:sweeps} varies these conditions, recording alignment, speech duration, and patient count. These are simulated workflows, not recordings of actual clinical encounters.

\vspace{-0.1cm}
\subsection{Extractors and comparisons}
\vspace{-0.1cm}
Table~\ref{tab:models} reports 11 frozen extractors from five families. Nine compact extractors use CPU inference through ONNX. We do not report a common runtime or latency comparison for all 11 models. TitaNet checkpoints include LibriSpeech in training, creating potential speaker overlap with this evaluation. Their results therefore require caution.

Baselines include AHC, PLDA adaptation \cite{castan2017}, and normalized maximum eigengap spectral clustering (NME-SC) \cite{park2020nme}, with raw or centered embeddings. Label-free methods do not use visitor identity labels for threshold selection. Oracle-informed comparisons receive a label-tuned threshold or the true person count and serve as privileged reference points, not deployable alternatives. ``Best baseline'' denotes the best reported score within each category for that extractor, rather than a single baseline selected across all extractors.

\vspace{-0.1cm}
\subsection{Metrics}
\vspace{-0.1cm}
We report visit-level B$^3$ precision, recall, and F1 \cite{bagga1998bcubed}, scaled by 100, along with the percentage of patients whose itinerary is recovered exactly. Table~\ref{tab:models} also reports the percentage of single-visit patients incorrectly merged with another identity. We report 95\% $t$-intervals over the 20 simulated days. Method differences use unpaired standard errors (conservative because all methods are evaluated on the same days).

\begin{table}[t]
\centering
\caption{Methods on the default simulated day (ECAPA-TDNN, 20 days): B$^3$-F1, precision, recall ($\times$100), and patients whose itinerary is recovered exactly (\%).}
\vspace{0.3cm}
\label{tab:methods}
\fontsize{9}{10.5}\selectfont
\setlength{\tabcolsep}{3pt}
\renewcommand{\arraystretch}{1.08}
\begin{tabularx}{\columnwidth}{@{}Yrrrr@{}}
\toprule
Method & F1 & P & R & Itin. \\
\midrule
\multicolumn{5}{@{}l}{\textit{Oracle-informed baselines}} \\
AHC, raw, oracle thr. & 72.4 & 80 & 66 & 31 \\
AHC, centered, oracle thr. & 87.7 & 91 & 84 & 61 \\
NME-SC, centered, true count & 76.4 & 76 & 77 & 32 \\
PLDA, centered, oracle thr. & 81.7 & 89 & 76 & 48 \\
\multicolumn{5}{@{}l}{\textit{Label-free}} \\
NME-SC, centered, auto & 57.6 & 59 & 84 & 27 \\
PLDA adapted, raw & 68.0 & 89 & 55 & 25 \\
PLDA adapted, centered & 76.4 & 97 & 63 & 41 \\
Ours, concurrency only & 68.3 & 96 & 57 & 35 \\
Ours, same-room rule everywhere & 78.2 & 89 & 70 & 41 \\
\textbf{Ours, room-type-aware} & 86.6 & 89 & 85 & 57 \\
\enspace + leakage guard & 86.9 & 89 & 86 & 58 \\
\enspace yesterday's statistics & 86.3 & 92 & 82 & 61 \\
\enspace online (causal greedy) & 82.2 & 87 & 79 & 50 \\
\bottomrule
\end{tabularx}
\end{table}

\section{Results}
\label{sec:results}

\vspace{-0.1cm}
\subsection{Default day and extractor dependence}
\vspace{-0.1cm}
Across all 11 extractors, the proposed method improves on the best label-free baseline by 5.3-14.6 B$^3$-F1 points (median paired gain: 8.3; Table~\ref{tab:models}); every gain is significant, and the largest (TitaNet) carry the training-overlap caveat of Section~\ref{sec:setup}. The medians of the baseline and proposed scores are 79.2 and 86.6, respectively; their difference is not the median paired gain. The median deficit to the best oracle-informed baseline is 1.9 points. The method exceeds that reference for two extractors, though neither difference is significant; it is statistically indistinguishable from it for ECAPA-TDNN and significantly below it for the remaining eight.

With ECAPA-TDNN, the method scores 86.6 versus 76.4 for the strongest label-free baseline and 87.7 for centered AHC with an oracle threshold (Table~\ref{tab:methods}). Exact itinerary recovery rises from 41\% for the label-free centered PLDA baseline to 57\%. Larger or more broadly trained extractors still help: w2v-BERT reaches 93.1 F1 and 76\% itinerary recovery, while the broadly trained ReDimNet2-B3 reaches 92.1 and 73\%. The 1.1M-parameter ReDimNet2-B0 improves from 79.2 to 84.5, but does not surpass the strongest unmodified extractor. Structure complements representation quality rather than removing its effect.

\vspace{-0.1cm}
\subsection{Constraints and operational variants}
\vspace{-0.1cm}
Applying the same-room rule to every room reduces ECAPA-TDNN F1 from 86.6 to 78.2, consistent with erroneous exclusions when patients return. The concurrency-only variant reaches 68.3; it also changes threshold calibration, so this difference is not a clean estimate of the room constraint's isolated contribution. The leakage guard yields 86.9, only 0.3 points above the default, whereas using the preceding day's statistics gives 86.3. Both differences lie within the 95\% intervals of the default. Causal greedy linking scores 82.2, a 4.4-point reduction relative to the offline method.

\vspace{-0.1cm}
\subsection{Sensitivity and failure modes}
\vspace{-0.1cm}
The proposed system beats the best label-free baseline for all 11 extractors in 13 of the 14 reported scenarios (Table~\ref{tab:sweeps}). Higher companion, return, silent-visit, leakage, and walk-in rates retain the advantage. A one-slot clock offset lowers the median score from 86.6 to 82.5, and the gain remains significant for only 5 of 11 extractors; reducing speech to 3\,s lowers the median to 81.5. The easier recorder mix increases the median to 91.6. Preprocessing is consequential. Without media rejection, the median drops to 80.2; it remains above the label-free baseline of 77.2, significantly so for 10 of 11 extractors. Without cross-room duplicate removal, it falls to 75.6 versus 80.1 for the baseline, and only 2 of 11 extractors retain an advantage. Acoustic duplication can turn presumed negative concurrent pairs into same-speaker observations, undermining both constraints and calibration. Aggregate clustering quality also hides identity risk. Across extractors, 35-72\% of single-visit patients are incorrectly merged. Even the strongest extractor therefore leaves substantial false-linking risk. A practical system would need explicit rejection or abstention criteria and evaluation of their coverage-error trade-off before links could inform clinical records.

\begin{table}[t]
\centering
\caption{Sensitivity: median B$^3$-F1 across 11 extractors, and wins over the best label-free baseline (sig.: wins whose 95\% interval excludes zero). LF and Oracle follow Table~\ref{tab:models}; default rates are given in Section~\ref{sec:setup}. Each row changes the named condition.}
\vspace{0.2cm}
\label{tab:sweeps}
\fontsize{9}{10.5}\selectfont
\setlength{\tabcolsep}{3pt}
\renewcommand{\arraystretch}{1.08}
\begin{tabularx}{\columnwidth}{@{}Yrrrr@{}}
\toprule
Scenario & LF & \textbf{Ours} & Oracle & Wins (sig.) \\
\midrule
Default day & 79.2 & \textbf{86.6} & 88.0 & 11/11 (11) \\
Companions: 95\% & 78.0 & \textbf{85.9} & 86.4 & 11/11 (11) \\
All lab patients return & 79.7 & \textbf{87.1} & 87.9 & 11/11 (11) \\
Companion speaks instead: 50\% & 79.0 & \textbf{87.6} & 88.4 & 11/11 (11) \\
Silent visits: 40\% & 78.4 & \textbf{86.5} & 87.7 & 11/11 (11) \\
Cross-room leakage: 30\% & 78.2 & \textbf{87.0} & 88.2 & 11/11 (11) \\
TV in every slot & 80.1 & \textbf{86.2} & 88.3 & 11/11 (11) \\
Clock offset: one slot & 79.1 & \textbf{82.5} & 87.4 & 11/11 (5) \\
55 patients/day & 78.4 & \textbf{87.2} & 87.2 & 11/11 (11) \\
3\,s per visit & 71.5 & \textbf{81.5} & 83.1 & 11/11 (11) \\
Walk-ins: 40\% & 81.2 & \textbf{87.8} & 87.7 & 11/11 (11) \\
Easy recorder mix & 87.6 & \textbf{91.6} & 93.6 & 11/11 (11) \\
No media (TV) rejection & 77.2 & \textbf{80.2} & 85.5 & 11/11 (10) \\
No cross-room duplicate removal & 80.1 & 75.6 & 87.8 & 2/11 (2) \\
\bottomrule
\end{tabularx}
\end{table}

\section{Limitations and Ethics}
\label{sec:limits}
\vspace{-0.2cm}
The evaluation uses simulated workflows and replayed read speech, with simulated phone processing. It does not establish performance on spontaneous clinical dialogue, multilingual conversations, or end-to-end diarization. The method assumes identified staff, usable visitor segments, aligned visit timing, appropriate room policies, and effective duplicate and media handling. Staff calibration also need not match the visitor score distribution. Checkpoint training overlap, export differences, and the limited number of simulated days limit cross-model conclusions. 
Only public speech corpora are used in the reported experiments and no clinical recordings are presented. Future clinical validation requires appropriate consent, data governance, and ethics review. Local embedding extraction may reduce raw-audio transfer, but embeddings and linked itineraries remain sensitive identity information.

\section{Conclusion}
\vspace{-0.2cm}
Spatiotemporal structure improves cross-device speaker linking across 11 extractors without visitor enrollment or visitor identity labels. On the default simulated day, gains over label-free baselines are 5.3-14.6 B$^3$-F1 points, all significant at the 95\% level. Room-specific constraints outperform a universal same-room rule, while duplicate-removal failures and frequent singleton false merges define the method's current limits. Real conversational recordings and explicit identity rejection are essential next steps.

\small
\bibliographystyle{IEEEbib}
\bibliography{paper}
\end{document}